# Understanding and reducing deleterious defects in metastable alloy GaAsBi

Guangfu Luo,[1] Shujiang Yang,[1] Glen R. Jenness,[2] Zhewen Song,[1] Thomas F. Kuech,[3] Dane Morgan,[1]

[1]*Department of Materials Science and Engineering, University of Wisconsin-Madison, Madison, Wisconsin 53706, USA*

[2]*Department of Chemistry, University of Wisconsin-Madison, Madison, Wisconsin 53706, USA*

[3]*Department of Chemical and Biological Engineering, University of Wisconsin-Madison, Madison, Wisconsin 53706, USA*

Correspondence: Professor Dane Morgan, Room 244, Department of Materials Science and Engineering, 1509 University Avenue, Madison, Wisconsin 53706, United States of America

E-mail: ddmorgan@wisc.edu, Telephone: +1-608-265-5879

Technological applications of novel metastable materials are frequently inhibited by abundant defects residing in these materials. Using first-principles methods we investigate the point defect thermodynamics and phase segregation in the technologically-important metastable alloy GaAsBi. Our calculations predict defect energy levels in good agreement with abundant previous experiments and clarify the defect structures giving rise to these levels. We find that vacancies in some charge states become metastable or unstable with respect to antisite formation, and this instability is a general characteristic of zincblende semiconductors with small ionicity. The dominant point defects degrading electronical and optical performances are predicted to be $As_{Ga}$, $Bi_{Ga}$, $Bi_{Ga}+Bi_{As}$, $As_{Ga}+Bi_{As}$, $V_{Ga}$ and $V_{Ga}+Bi_{As}$, of which the first-four and second-two defects are minority-electron and minority-hole traps, respectively. $V_{Ga}$ is also found to play a critical role in controlling the metastable Bi supersaturation through mediating Bi diffusion and clustering. To reduce the influences of these deleterious defects, we suggest shifting the growth away from As-rich condition and/or using hydrogen passivation to reduce the minority-carrier traps. We expect this work to aid in the applications of GaAsBi to novel electronic and optoelectronic devices, and shine a light on controlling the deleterious defects in other metastable materials.

INTRODUCTION

The metastable alloy GaAsBi has been intensely studied since its first successful fabrication by metalorganic vapor phase epitaxy (MOVPE) in 1998[1] and by molecular beam epitaxy (MBE) in 2003,[2] due to three key advantages over other GaAs-based compounds. First, Bi atoms reduce the band gap much more effectively than other alloying elements, such as In (~83 meV per Bi% versus ~15 meV per In%).[3] Second, Bi incorporation dramatically decreases the temperature dependence of electronic properties. For instance, the temperature coefficient of band gap of a GaAsBi film with 2.6% Bi is only 1/3 of that of GaAs near room temperature.[4] This property can lead to the realization of temperature-insensitive devices,[5] as confirmed in the study of GaAsBi-based laser diodes,[6] and benefits a number of technologies. Third, Bi atoms strongly increase the spin-orbit (SO) splitting of GaAsBi, and with Bi contents near 10% it is expected to totally suspend one type of Auger recombination,[3, 5] which induces the serious "efficiency droop" in high-energy light-emitting diodes and laser diodes. These advantages suggest that GaAsBi would provide a major improvement in many semiconductor devices as long as high-quality samples with significant Bi contents can be obtained.

However, Bi has a very low solubility in GaAs; incorporation of a large Bi content (~10%) with low defects and without phase segregation is especially challenging.[7] In the usual growth temperature range of GaAs, 500–600 °C, the Bi incorporation is nearly zero, confirmed by both experiments[8] and theoretical claculations.[9] To enhance the Bi incorporation, growth temperatures as low as 300–400 °C are widely used to slow down the growth dynamical processes that lead to low Bi incorporation. However, the low growth temperature also induces numerous defects that cause short carrier lifetimes and low photoluminescence intensity.[10, 11] Additionally, a previous theoretical study[12] has shown that the dynamical processes associated with As replacing Bi, which limits Bi incorporation, are still very quick at low temperature, e.g. <$10^{-6}$ s at 320 °C. To reduce the defects, thermal annealing has been used, but the incorporated Bi atoms often segregate and form Bi-rich clusters.[13] These defect issues have become a bottleneck of the successful applications of GaAsBi. For instance, the GaAsBi–based laser diodes up to date require about an order of magnitude higher threshold current density (2–10 kA/cm$^2$)[6, 14-16] than usual InGaAs–based laser diodes (0.2–0.5 kA/cm$^2$).

In the current work, we commit to answer several essential questions relevant to the defects in GaAsBi. What are the primary intrinsic point defects and how do they interact to form pair defects under equilibrium condition? How do the defects affect minority-carriers? What defects mediate the formation of Bi-rich clusters during thermal annealing? Finally, how can one reduce the deleterious effects of the dominant defects? Our major results are based on the thermodynamic properties of defects, which are expected to give a useful guide to many aspects of defect behavior, even under non-equilibrium conditions. Additionally, we will qualitatively discuss the non-equilibrium effects when the equilibrium results become invalid.

MODELS AND METHODS

In this paper, we examine six point defects: $V_{Ga}$, $V_{As}$, $As_{Ga}$, $Ga_{As}$, $Bi_{As}$ and $Bi_{Ga}$. The interstitial defects of As and Ga are known for their high formation energies in GaAs[17, 18] and are excluded, so is the Bi interstitial. To understand how the dominant point defects interact with each other, we further examine thirteen pair defects, each consisting of two nearest-neighbor point defects. Motived by the findings of pair defects, we examine eight clusters involving Bi defects with and without vacancies to explore the formation of Bi-rich clusters. Finally, we investigate eight defects involving H to examine the effects of hydrogen passivation.

We carry out *ab initio* calculations using density functional theory (DFT) as implemented in the Vienna *ab initio* Simulation Package.[19] An energy cutoff of 400 eV is set to the plane-wave basis set and the following projector-augmented wave potentials are utilized: Ga_GW($4s^24p^1$) for Ga, As_GW($4s^24p^3$) for As, Bi_d_GW($5d^{10}6s^26p^3$) for Bi, and H_GW($1s^1$) for H. These GW potentials give generally better properties for excited electronic states than the standard DFT potentials and are therefore appropriate for our defect calculations involving charge transfer to excited states. The HSE06[20] hybrid functional is used to correctly describe the band gap. The predicted band gap of GaAs bulk using HSE06 is 1.38 eV, consistent with the experimental value of 1.42 eV at 300 K.[21] To describe the strong special relativity effects in Bi atom, the SO coupling is included for the calculations involving Bi. The predicted SO splitting at the Γ point is 0.39 eV for bulk GaAs, in excellent agreement with the experimental value of 0.34 eV at 300 K.[21] In Supplementary Fig. S1, we demonstrate the importance of using both HSE06 and SO coupling by showing the notable differences in defect formation energies using different methods. In our HSE06 calculations, the supercell size is 2 × 2 × 2, with a volume of $11.2^3$ Å$^3$, and the **k**-point sampling is a 4 × 4 × 4 Monkhorst-Pack grid. The *ab initio* method proposed by Freysoldt, Neugebauer and Van de Walle (FNV)[22] is adopted to remove the image charge interaction and adjust the potential alignment between the perfect and defected structures. Our tests suggest that the error of defect formation energy after the FNV correction is limited to about 0.1 eV (Supplementary Fig. S2). Symmetry is broken to allow possible Jahn–Teller distortion around the defects. To automatically prepare the initial structures, manage the workflow and post-process the results, the high-throughput computational tool, MAterials Simulation Toolkit (MAST)[23] is used. Note that three major factors,[24] namely, correct description of band gap, proper removal of the image charge interaction and proper potential alignment, are critical to obtain reliable defect formation energies. Inadequate treatment of the three factors in previous studies of GaAs bulk resulted in quite scattered data (Supplementary Fig. S3).

The defect formation energy $E_f$ of a defect $A_B$ (atom $A$ on host site $B$) with a charge state $q$ is defined as Eqn. 1,[24]

$$E_f(A_B^q, E_F) \equiv E_{tot}(A_B^q) + E_{FNV}(A_B^q) - E_{tot}(\text{GaAs}) + \mu_B - \mu_A + q\left[E_{VBM}(\text{GaAs}) + E_F\right] \quad (1)$$

where $E_{tot}(A_B^q)$ and $E_{tot}$(GaAs) are the total internal energy of the system with defect $A_B^q$ and perfect GaAs bulk, respectively; $E_{FNV}$, $E_{VBM}$ and $E_F$ is the FNV correction, valence band maximum (VBM) energy and Fermi energy relative to VBM energy, respectively; chemical potentials $\mu_X$ for $X$ = Bi, As and Ga at different conditions are summarized in Table 1. The lowest defect formation energy of a defect, $E_f^{min}$, is determined by the minimum value of different charge states,

$$E_f^{min}(A_B, E_F) \equiv \text{Min}\{E_f(A_B^q, E_F)\}. \quad (2)$$

Defect energy levels correspond to the $E_F$ where the slope of $E_f^{min}$ changes and are independent of the choice of chemical potentials. To explore the interaction between different defects, we define the binding energy, $E_b$, of a complex defect by Eqn. 3.

$$E_b(A_B + C_D, E_F) \equiv E_f^{min}(A_B + C_D, E_F) - E_f^{min}(A_B, E_F) - E_f^{min}(C_D, E_F), \quad (3)$$

where the complex defect $A_B+C_D$ consist of defects $A_B$ and $C_D$ bound together.

**Table 1.** Chemical potential $\mu$ (eV) of Bi, As and Ga under As-rich, Ga-rich and intermediate conditions. $\mu_{Bi}$ is the total energy per atom of the rhombohedral Bi metal, since a significant number of Bi layers or droplets exist on the film surface under usual growth conditions.[12] $\mu_{As}$ ($\mu_{Ga}$) under the As-rich (Ga-rich) condition equals the total energy per atom of the hexagonal As (orthorhombic Ga) bulk. Sum of $\mu_{Ga}$ and $\mu_{As}$ under the same condition equals the total energy per chemical formula of GaAs bulk. The absolute difference of $\mu_{Ga}$ or $\mu_{As}$ between the As- and Ga-rich conditions equals the standard formation enthalpy of GaAs, and our theoretical value of 0.95 eV is in good agreement with a recent experimental measurement of 0.91 eV.[25] $\mu$ of the intermediate condition is defined as the average value of the As- and Ga-rich conditions.

|  | As-rich | Ga-rich | Intermediate |
|---|---|---|---|
| $\mu_{Bi}$ | −6.00 | −6.00 | −6.00 |
| $\mu_{As}$ | −5.93 | −6.88 | −6.40 |
| $\mu_{Ga}$ | −4.60 | −3.65 | −4.13 |

## RESULTS AND DISCUSSIONS

**Metastability of defects involving vacancy.** Our first finding is that in certain charge states, where the formation energies are generally very high, the point and pair defects involving vacancies become metastable or unstable. This instability is that the cation and anion vacancies can transform between each other with the assistance of an antisite defect, a phenomenon first discovered theoretically for isolated $V_{As}$ and $V_{Ga}$ over 30 years ago.[26, 27] Table 2 shows that $V_{As}$ can change to $V_{Ga}+Ga_{As}$ at 2− and 3− charge states, and $V_{Ga}$ can change to $V_{As}+As_{Ga}$ at 1+, 2+ and 3+ charge states. As we will show later in Fig. 1a, these charge states correspond to $V_{As}$ and $V_{Ga}$ with high formation energies relative to the other charge states. Because $V_{As}$ and $V_{Ga}$ at the above charge states are metastable, one needs to directly examine the "swapped" structures to reveal their greater stability, which possibly explains the neglect of these swapped structures in recent studies of GaAs.[18, 28, 29] The anion (cation) vacancy in the pair defects $V_{As}+Bi_{As}$, $V_{Ga}+Bi_{Ga}$, $V_{Ga}+As_{Ga}$ and $V_{Ga}+Ga_{As}$ at certain charge states can all undergo similar structural changes to a more stable form with a cation (anion) vacancy and an antisite replacing the original vacancy. For the three other pair defects, $V_{As}+Bi_{Ga}$, $V_{As}+As_{Ga}$ and $V_{Ga}+Ga_{As}$, they can change to a more stable form at certain charge states by hopping the Bi, As and Ga to the neighboring vacancy, respectively.

**Table 2.** Structural changes of point and pair defects involving vacancy. $Q_f$ is the total charge state of the initial and resulting defects when the forward change is favorable. The backward change is favorable at all the other charge states in the range of [3−, 3+].

| point defect | | pair defect | |
|---|---|---|---|
| reaction | $Q_f$ | reaction | $Q_f$ |
|  |  | $V_{As}+Bi_{As} \leftrightarrow V_{Ga}+Ga_{As}+Bi_{As}$ | 2−, 3− |
| $V_{As} \leftrightarrow V_{Ga}+Ga_{As}$ | 2−, 3− | $V_{As}+Bi_{Ga} \leftrightarrow V_{Ga}+Bi_{As}$ | 3− − 0 |
|  |  | $V_{As}+As_{Ga} \leftrightarrow V_{Ga}$ | 3− − 0 |
|  |  | $V_{Ga}+Bi_{Ga} \leftrightarrow V_{As}+As_{Ga}+Bi_{Ga}$ | 1+ − 3+ |
| $V_{Ga} \leftrightarrow V_{As}+As_{Ga}$ | 1+ − 3+ | $V_{Ga}+As_{Ga} \leftrightarrow V_{As}+As_{Ga}+As_{Ga}$ | 1+ − 3+ |
|  |  | $V_{Ga}+Bi_{As} \leftrightarrow V_{As}+Bi_{Ga}$ | 1+ − 3+ |
|  |  | $V_{Ga}+Ga_{As} \leftrightarrow V_{As}$ | 1− − 3+ |

Experimental observation of the aforementioned metastability can be challenging, because it occurs under conditions where the defects have relatively high formation energies and therefore low concentrations, e.g. n-type doping condition for $V_{As}$ and p-type doping condition for $V_{Ga}$. Nevertheless, a possible experimental observation of the change $V_{As}+As_{Ga} \rightarrow V_{Ga}$ was suggested in an early deep-level transient spectroscopy (DLTS) study,[30] which claimed that the signal decrease of $V_{As}+As_{Ga}$ was accompanied by the signal increase of $V_{Ga}$ during the treatment of n-type GaAs using ultrasonic vibration. To explore the generality of vacancy metastability in compound semiconductors, we examine the anion vacancy of twelve zincblende binary compounds, including ZnO, SiC and ten III-V compounds. Results indicate that the metastability is quite general but also closely related to ionicity: all the examined compounds with ionicity less than 0.5 undergo similar change as $V_{As}$ in GaAs (Supplementary Table S1).

**Formation energies of point and pair defects.** Figure 1a shows the defect formation energy of the point defects under As-rich condition, which is widely used by providing excessive As reactant in the growth chamber. The influence of other chemical potentials on the defect formation energy will be discussed at the end of this paper. The aforementioned metastability significantly changes the properties of $V_{Ga}$ and $V_{As}$ under the p- and n-type doping conditions, respectively. The dominant defects are $As_{Ga}^{2+}$, $Bi_{Ga}^{2+}$ and $Bi_{As}^{0}$ in the p-type films, while $V_{Ga}^{3-}$ and $Bi_{As}^{0}$ in the n-type films. In contrast, $V_{As}$ and $Ga_{As}$ have much greater formation energy and are expected to have insignificant concentrations at or near thermal equilibrium. Note that although $Bi_{As}$ and $Bi_{Ga}$ do not have the lowest defect formation energy under most doping conditions, their non-equilibrium contents are expected to be dominate because of the significant (typically >1%) overall Bi contents in most GaAsBi films.

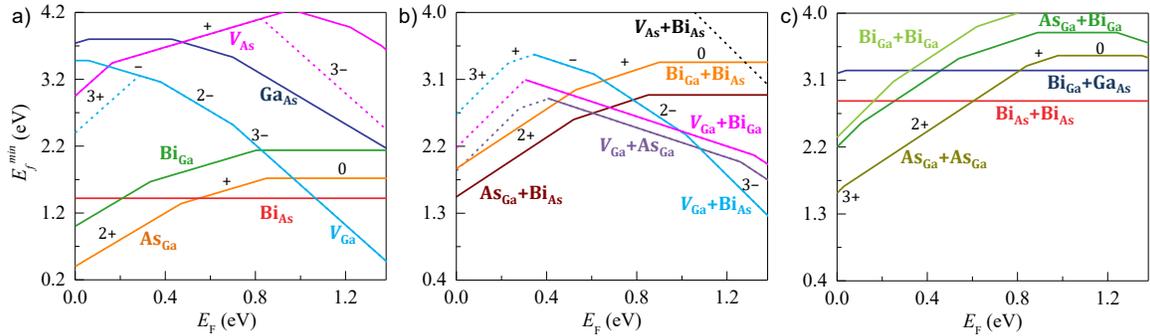

**Figure 1.** Lowest defect formation energy versus Fermi energy for (**a**) point defects and (**b**, **c**) pair defects under the As-rich condition. The dashed branches correspond to the stable swapped structures as listed in Table 2; in panel **a**, the metastable states of $V_{Ga}$ and $V_{As}$ are shown for reference. The results of pair defects are plotted in two panels for easy visualization.

One interesting result is that under the condition $E_F < 0.21$ eV, the equilibrium content of $Bi_{Ga}^{2+}$ is predicted to be substantially more than $Bi_{As}^{0}$. This prediction suggests that the usual assumption, namely that incorporated Bi primarily substitutes As due to their chemical similarity, does not hold under certain growth conditions. Previous growth of Bi-doped GaAs using the liquid encapsulated Czochralski method indeed found a significant amount (~10%) of Bi in the form of $Bi_{Ga}$.[31] It is worthy pointing out that another study[32] estimated the maximum ratio of $Bi_{Ga}$ to be only ~5% of the total incorporated Bi atoms. That estimation was based on a fitting of the theoretical structural expansion around $Bi_{As}$ and $Bi_{Ga}$ with experimental X-ray absorption spectra, and a fundamental parameter utilized was that the Bi-As bonds around $Bi_{Ga}$ are 0.15 Å longer than the Bi-Ga bonds around $Bi_{As}$. However, our calculations indicate that the aforementioned bond length difference varies dramatically with the charge state of $Bi_{Ga}$. Namely, the Bi-As bonds around $Bi_{Ga}^{2+}$, $Bi_{Ga}^{+}$ and $Bi_{Ga}^{0}$, are 0.05, 0.11 and 0.15 Å longer than the Bi-Ga bonds around $Bi_{As}^{0}$, respectively. The previous study therefore applied bond lengths appropriate only for the structures of $Bi_{Ga}^{0}$ and $Bi_{As}^{0}$. If $Bi_{Ga}^{2+}$ in the p-type films were considered, the fitting would yield a significantly higher fraction of $Bi_{Ga}$. Because $Bi_{Ga}$ possesses deep energy levels, which are significant carrier traps as discussed below, but isolated $Bi_{As}$ does not, further experimental verification of the Bi substitution sites under different growth conditions is a critical need.

To verify our results, we compare the predicted energy levels of the point defects with available experimental data in Fig. 2. We find that the average deviation between our predictions and the experimental values is 0.06 eV, which is encouraging given the much larger deviation found in previous theoretical studies (Supplementary Fig. S3). For the most extensively studied defect $As_{Ga}$, which corresponds to the EL2 peak in DLTS experiments, our predicted positions are 0.46 and 0.84 eV for the 2+/+ and +/0 energy levels, which agree reasonably well with the average experimental values of 0.49 and 0.72 eV, respectively.

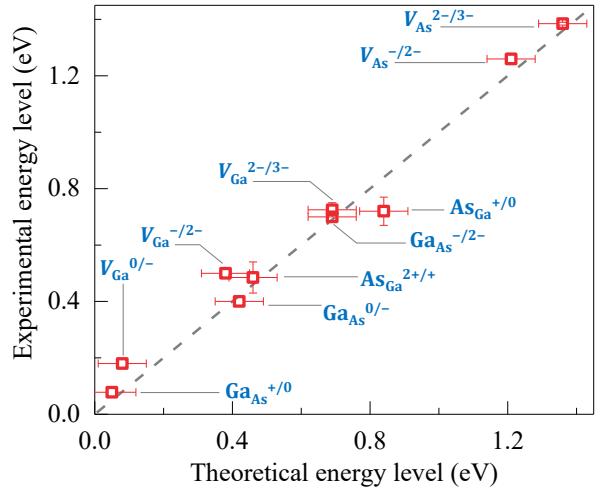

**Figure 2.** Comparison of defect energy levels between experiments and predictions by this work. The error bars represent the range of different values of the experimental data ($V_{Ga}$ from Ref. 33, 34, $V_{As}$ from Ref. 35, 36, $As_{Ga}$ from Ref. 31, 37-39, and $Ga_{As}$ from Ref. 40-42) or the supercell convergence errors of the theoretical data.

Additionally, our results question the accuracy of early reports on the energy levels of $V_{As}$ and $Bi_{Ga}$. In Fig. 2, the two energy levels of $V_{As}$ near the conduction band minimum were identified as +/0 and 0/− by Saarinen et al.[35] but 0/− and −/2− by Loualiche et al.[36], both of which are different from our prediction of −/2− and 2−/3−. Kunzer at al.[31] conjectured that the energy level of $Bi_{Ga}^{2+/+}$ and $Bi_{Ga}^{+/0}$ lies in the ranges of 0.72 – 0.77 eV and 0.92 – 1.07 eV, respectively, which deviate significantly from our predictions of 0.33 and 0.80 eV.

Figure 1b and 1c show the defect formation energies of the pair defects. We find that the dominant pair defects originate from the combinations of dominant point defects ($As_{Ga}$, $Bi_{As}$, $Bi_{Ga}$ and $V_{Ga}$). Under p-type condition, the dominant pair defect is $(As_{Ga}+Bi_{As})^{2+}$. Though $(As_{Ga}+As_{Ga})^{2+,3+}$ also has relatively low formation energies under p-type condition, later we will show that two $As_{Ga}$ defects have a repulsive interaction under such condition and thus the pair is unlikely to form. Under n-type condition, the dominant pair defects are $(V_{Ga}+Bi_{As})^{3-}$ and $(V_{Ga}+As_{Ga})^{1-,2-}$, both involving $V_{Ga}$. It should be noted that non-equilibrium growth of most GaAsBi films leads to much higher total Bi contents than the equilibrium values. To understand the potential influences of non-equilibrium conditions on Bi defects, we calculate the defect formation energies under the condition of the total Bi content being fixed to 1%, which is established by allowing the Bi chemical potential $\mu_{Bi}$ to vary with $E_F$ as necessary (see Supplementary Information V). We find that five Bi defects: $Bi_{As}$, $Bi_{Ga}$, $Bi_{Ga}+Bi_{As}$, $Bi_{As}+Bi_{As}$ and $V_{Ga}+Bi_{As}$ would have significant concentrations in certain regions of $E_F$ (Supplementary Fig. S4).

With the above understanding, we can now clarify the nature of the majority-carrier traps reported by recent DLTS experiments on GaAsBi.[43] As listed in Table 3, the three experimental majority-electron traps in n-type films can be assigned to our predicted primary defect levels of $(V_{Ga}+Bi_{As})^{2-/3-}$, $V_{Ga}^{2-/3-}$ and $(V_{Ga}+Bi_{As})^{-/2-}$ with an average deviation of ~0.12 eV between experiments and this work. For the six majority-hole traps in p-type films, the three highest energy levels can be assigned to {$Bi_{Ga}^{+/0}$, $As_{Ga}^{+/0}$, $(As_{Ga}+Bi_{As})^{+/0}$, $(Bi_{Ga}+Bi_{As})^{+/0}$}, {$As_{Ga}^{2+/+}$, $(As_{Ga}+Bi_{As})^{2+/+}$, $(Bi_{Ga}+Bi_{As})^{2+/+}$} and $Bi_{Ga}^{2+/+}$, with an average deviation of ~0.04 eV between experiments and this work. As we will show later, the lowest defect level at 0.08 eV could be assigned to Bi-rich clusters with vacancies, such as $V_{As}+4Bi_{As}$ and $V_{Ga}+4Bi_{As}$. The nature of energy levels at 0.12 and 0.17 eV is unclear and could be ascribed to unexplored complex defects and/or impurities in the films.

**Table 3.** Comparison between previous experiments[43] and this work for primary majority-electron traps in n-type GaAsBi films and majority-hole traps in p-type ones. Energy levels of majority-electron traps are relative values below conduction band minimum and majority-holes traps above VBM. Defect origins predicted by this work are listed for corresponding energy levels.

|  | Experimental energy (eV) | This work | |
|---|---|---|---|
|  |  | Energy (eV) | Defect origin |
| Majority-electron trap | 0.23–0.28 | 0.36 | $(V_{Ga}+Bi_{As})^{2-/3-}$ |
|  | 0.56–0.61 | 0.69 | $V_{Ga}^{2-/3-}$ |
|  | 0.60–0.67 | 0.78 | $(V_{Ga}+Bi_{As})^{-/2-}$ |
| Majority-hole trap | 0.87–0.88 | 0.80, 0.84, 0.84, 0.89 | $Bi_{Ga}^{+/0}$, $As_{Ga}^{+/0}$, $(As_{Ga}+Bi_{As})^{+/0}$, $(Bi_{Ga}+Bi_{As})^{+/0}$ |
|  | 0.50–0.53 | 0.46, 0.52, 0.52 | $As_{Ga}^{2+/+}$, $(As_{Ga}+Bi_{As})^{2+/+}$, $(Bi_{Ga}+Bi_{As})^{2+/+}$ |
|  | 0.27–0.30 | 0.33 | $Bi_{Ga}^{2+/+}$ |
|  | 0.17 | — | — |
|  | 0.12 | — | — |
|  | 0.08 | 0.02–0.08 | $V_{As}+nBi_{As}$, $V_{Ga}+nBi_{As}$ |

**Influences of defect energy levels on minority-carriers.** To identify the most deleterious defect energy levels to minority-carriers, which are critical to the electron-hole recombination in optoelectronic devices, we visualize the strength of minority-carrier trapping as a function of Fermi energy for each defect energy level in Fig. 3. The trapping rate[44] of minority-electron (minority-hole) for a given defect energy level, $r_{e\text{-}trap}$ ($r_{h\text{-}trap}$), is determined by the product of three factors, namely, defect concentration, $c_d$, non-occupancy fraction of the defect energy level for electron (hole), $f_{e,l}$ ($f_{h,l}$), and capture cross-section of electron (hole), $\sigma_e$ ($\sigma_h$):

$$r_{e-trap} = c_d f_{e,l} \sigma_e = e^{-E_f/k_BT}\left(1 - \frac{1}{1+e^{(E_l-E_F)/k_BT}}\right)\sigma_e$$

$$r_{h-trap} = c_d f_{h,l} \sigma_h = e^{-E_f/k_BT} \frac{1}{1+e^{(E_l-E_F)/k_BT}} \sigma_h. \quad (4)$$

Here $E_f$ is the defect formation energy, $c_d = e^{-E_f/k_BT}$ the defect concentration, and $E_l$ the position of defect energy level. Previous studies found that Coulomb interaction between the carriers and defects plays a critical role: the typical capture cross-section of repulsive centers, neutral centers and attractive centers are $10^{-8}$–$10^{-5}$, $10^{-1}$–$10^{1}$ and $10^{1}$–$10^{4}$ Å$^2$, respectively.[45] Here we approximate $\sigma_e$ and $\sigma_h$ to the intermediate values of the abovementioned ranges as written in Eqn. 5.

$$\sigma_e = \begin{cases} 5\times10^3, & q>0 \\ 5, & q=0 \\ 5\times10^{-6}, & q<0 \end{cases}, \quad \sigma_h = \begin{cases} 5\times10^3, & q<0 \\ 5, & q=0 \\ 5\times10^{-6}, & q>0 \end{cases} \quad (5)$$

where $q$ is charge state of defect. Tests show that the exponential prefactor dominates Eqn. 4 and even two orders of uncertainty of $\sigma_e$ and $\sigma_h$ results in minor changes to Fig. 3.

Figure 3 shows that the minority-electron traps are dominated by $As_{Ga}$, $Bi_{Ga}$, $As_{Ga}+Bi_{As}$ and $Bi_{Ga}+Bi_{As}$ under p-type doping condition, and minority-hole traps by $V_{Ga}$ and $V_{Ga}+Bi_{As}$ under n-type doping condition. Note that because $Bi_{Ga}+Bi_{As}$ is expected to have a significant concentration under non-equilibrium conditions (Supplementary Fig. S4) in spite of its high defect formation energy, we reduce $E_f$ by 0.43 eV per Bi atom for all Bi defects (an amplitude corresponding to half of the $\mu_{Bi}$ decrease at VBM in the aforementioned non-equilibrium estimation) to include $Bi_{Ga}+Bi_{As}$ in Fig. 3. A similar figure without this correction can be found in Supplementary Fig. S5. Since all the primary traps contain at least one of the unwanted point defects, $As_{Ga}$, $Bi_{Ga}$ and $V_{Ga}$, one should be able to effectively minimize the traps by efficient reduction of the three point defects.

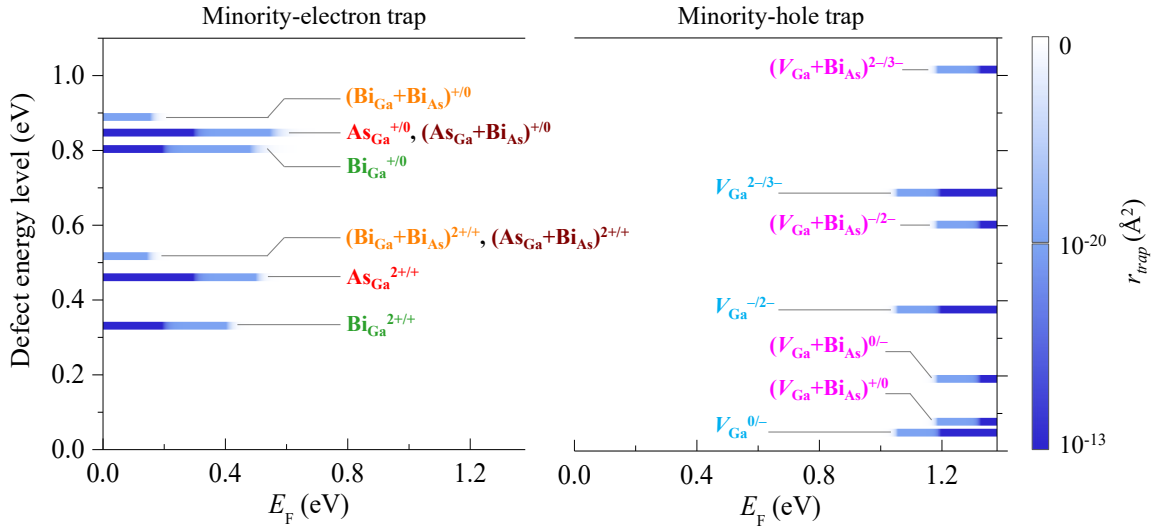

**Figure 3.** Trapping rate as a function of Fermi energy for (**Left**) minority-electron and (**Right**) minority-hole traps induced by the energy levels of all examined point and pair defects. Energy levels with extremely low trapping rate are invisible within the color mapping scheme. Temperature is set to 300 K.

**Binding energy of pair defects and formation of Bi-rich clusters.** To understand how point defects interact and form pair defects (important processes during thermal annealing), we plot the binding energy of pair defects in Fig. 4. We find that $Bi_{As}+Bi_{As}$ has slightly negative (-0.02 eV) binding energies in the whole range of Fermi energy. $Bi_{Ga}+Bi_{Ga}$, $As_{Ga}+As_{Ga}$ and $As_{Ga}+Bi_{Ga}$ have positive binding energies when $E_F$ is less than 0.29, 0.40 and 0.65 eV, respectively, and slightly negative binding energies otherwise. In view of their relatively high formation energies (Fig. 1c), the three pair defects are expected to form in only a small quantity and over a limited range of Fermi energy. By contrast, all the other pair defects, including those with high contents under equilibrium or non-equilibrium conditions ($As_{Ga}+Bi_{As}$, $Bi_{Ga}+Bi_{As}$ and $V_{Ga}+Bi_{As}$), possess negative binding energies in the entire range of Fermi energy. The pair defects involving Bi and vacancies have particularly low binding energies and thus the Bi-vacancy pairs likely provide the nuclei of the Bi-rich clusters during thermal annealing. Though $Bi_{Ga}+Ga_{As}$ also has low binding energy and therefore $Bi_{Ga}$ and $Ga_{As}$ do form pairs, its overall formation energy is over 3.1 eV (Fig. 1c), which means that its content should be low, even in the presence of a significant Bi content under non-equilibrium conditions (Supplementary Fig. S4).

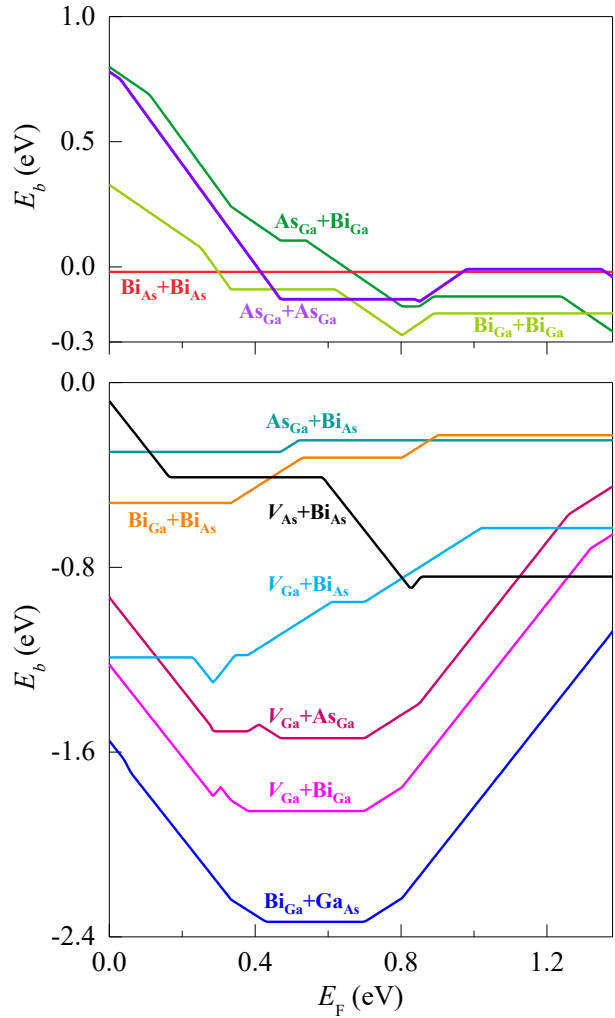

**Figure 4.** Binding energies of pair defects relative to isolated point defects. The plots are shown in two panels for easy visualization.

To further examine if a vacancy can bind multiple Bi defects together, we study relatively large clusters with and without a vacancy in the cluster center, namely, $n$Bi$_{As}$, $n$Bi$_{Ga}$, $V_{Ga}$+$n$Bi$_{As}$ and $V_{As}$+$n$Bi$_{As}$ ($n$ = 3 and 4). It turns out that pure Bi clusters have repulsive interactions in most regions of $E_F$, but a vacancy strongly binds the Bi defects together (Supplementary Fig. S6). Such behavior can be partially explained by the compensation of strain induced by Bi defects and vacancies. Our results support and extend the conclusion of a previous DFT study,[46] which reported the nucleus effect of $V_{Ga}^0$ in forming the Bi-rich clusters using the LDA functional. Besides strong binding strength, the formation of Bi-rich clusters with vacancies rather than pure Bi clusters is supported by another fact: $V_{As}$+3Bi$_{As}$, $V_{As}$+4Bi$_{As}$ and $V_{Ga}$+4Bi$_{As}$ possess energy levels 0.02–0.08 eV above VBM (Supplementary Fig. S6a and 6b), which agree with previous experimental observation of Bi-included localized states 0.05 – 0.09 eV above VBM.[47] However, all the examined pure Bi defects ($m$Bi$_{As}$ for $m$ = 1 – 4 and $n$Bi$_{Ga}$ for $n$ = 1 – 2) do not have similar energy levels.

Besides serving as nuclei of the Bi-rich clusters, vacancies are expected to play a critical role in assisting the diffusion of Bi during the cluster formation, as the very large size of Bi atoms relative to Ga and As makes it unlikely to be an interstitial diffuser. Unfortunately, complete and quantitative modeling of the defect-mediated Bi diffusion over all possible Fermi levels and external chemical potentials is very complex and is beyond the scope of this paper. We therefore focus on building a simplified model under As-rich and *n*-type doping conditions. If we make a reasonable assumption that Bi diffuses by a vacancy-mediated mechanism with nearest-neighbor hops on either the Ga or As sublattice, then the diffusions must involve the formation of pair defect $V_{Ga}$+Bi$_{Ga}$ or $V_{As}$+Bi$_{As}$. Because $V_{Ga}$+Bi$_{Ga}$ is significantly more stable than $V_{As}$+Bi$_{As}$ for all Fermi levels (Fig. 1b) and has significantly stronger binding strength than $V_{As}$+Bi$_{As}$ for almost all Fermi levels (Fig. 4), it is likely that the primary Bi diffusion path is through $V_{Ga}$-mediated diffusion of Bi$_{Ga}$. Additionally, Fig. 1b shows that $V_{Ga}$+Bi$_{Ga}$ is stable in the charge state 1- over a wide range of Fermi level. Therefore, we focus on Bi$_{Ga}$ diffusion through $V_{Ga}$-mediated hops with the whole system in 1- charge state; the diffusion of isolated $V_{Ga}$ is assumed to be in the dominant charge state 3-.

We calculate the $V_{Ga}$-mediated Bi$_{Ga}$ diffusion using the well-known 5-frequency model, as elaborated in Supplementary Information VIII. Six critical parameters for the model, namely, migration barriers of processes 0–4 and self-diffusivity of Ga, are found to be 2.16 eV, 2.47 eV, 1.79 eV, 2.54 eV, 1.98 eV and $4.3 \times 10^{17} e^{(6.78eV-3E_F)/(k_B T)}$ Å$^2$/sec, respectively. We find that in the typical annealing temperature range of 600–800 °C and annealing time range of 60–120 seconds, the diffusion length of Bi$_{Ga}$ is long enough to form Bi-rich clusters under n-type doping condition. For example, a 60-second annealing at 800 °C leads to a diffusion length of Bi$_{Ga}$ over 20 Å in the case of $E_F$ > 0.91 eV. In comparison, for a GaAsBi film with 1.0% to 3.0% uniformly distributed Bi atoms, the average distance between two neighboring Bi atoms is only about 16 to 11 Å. Therefore, $V_{Ga}$ is capable of assisting Bi$_{Ga}$ diffusion to form Bi-rich clusters in typical GaAsBi films with typical annealing temperature and time. This mechanism of Bi clustering is also supported by the fact that $V_{Ga}$ assists the diffusion of As$_{Ga}$ to form As-rich clusters during the thermal annealing of low temperature-grown GaAs.[48, 49] Because of the strong binding between $V_{Ga}$ and Bi$_{Ga}$, as well as the relatively fast $V_{Ga}$-mediated Bi$_{Ga}$ diffusion, we anticipate that the initial Bi-rich clusters consist of both Bi and $V_{Ga}$, and minimizing the content of $V_{Ga}$ is expected to be an effective way of reducing the formation of Bi-rich clusters under thermal annealing.

**Hydrogen passivation of defects.** To explore methods of reducing the deleterious influences of defects, we first examine the hydrogen passivation of the dominant defects, a technology widely used in the silicon industry to reduce carrier traps. Figure 5a shows that hydrogen can effectively passivate $V_{Ga}$: the number of deep energy levels of $V_{Ga}+n$H is reduced from three to zero with the number of hydrogen atoms, $n$, increasing from zero to three. Interestingly, adding a fourth hydrogen atom around $V_{Ga}$ does not induce extra defect energy level in the gap, which therefore allows a relatively wide range of hydrogen passivation condition without creating new gap states. Thermodynamically, the H-passivated $V_{Ga}$ are much more stable (over 0.53 eV per $V_{Ga}$) relative to the isolated $V_{Ga}$ and hydrogen interstitial $H_i$, as shown in Fig. 5c. Note that $H_i$ is known to be able to exist in GaAs bulk under common MOVPE conditions.[50] Therefore, hydrogen passivation is likely to be realized and could effectively remove the defect levels induced by $V_{Ga}$. An extra advantage is that $V_{Ga}$ is expected to diffuse significantly slower after passivation, which will reduce $Bi_{Ga}$ diffusion and consequently the formation of Bi-rich clusters. Because of the similarity between $V_{Ga}$ and $V_{Ga}+Bi_{As}$, it is likely that hydrogen passivation can also effectively remove the defect energy levels of $V_{Ga}+Bi_{As}$.

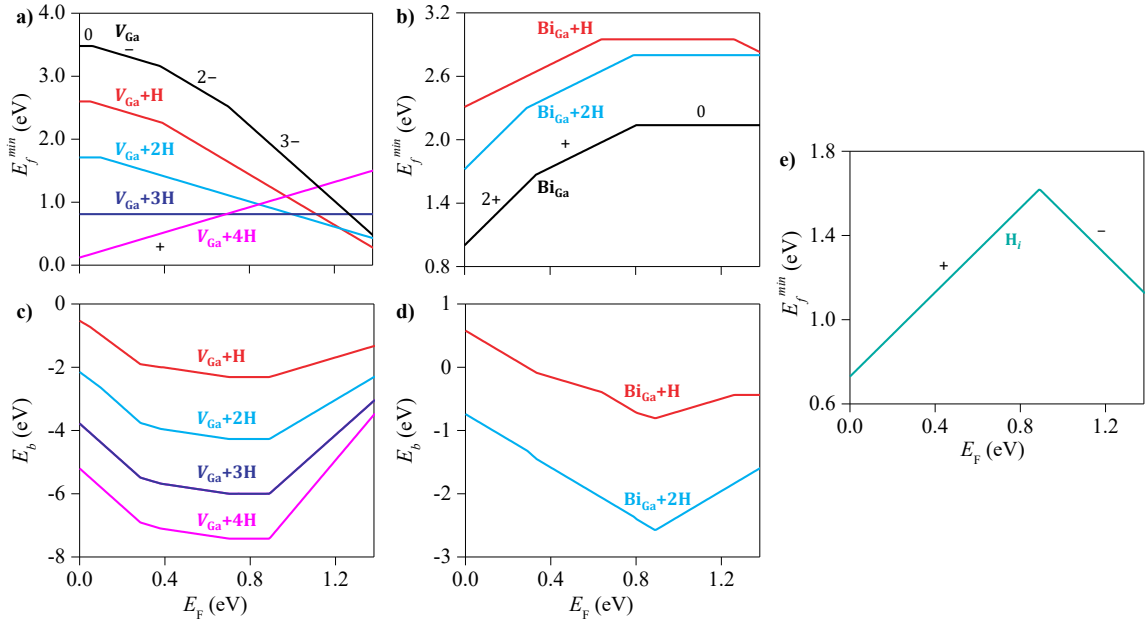

**Figure 5.** Lowest defect formation energies of (**a**) $V_{Ga}+n$H and (**b**) $Bi_{Ga}+n$H, and binding energies of (**c**) $V_{Ga}+n$H and (**d**) $Bi_{Ga}+n$H. (**e**) Lowest defect formation energies of $H_i$. The binding energy is relative to isolated $H_i$ and $V_{Ga}$ or $Bi_{Ga}$. Chemical potential of H is chosen to be half of the energy of a $H_2$ molecule (other choices of chemical potential only shift the curve in panel e up or down). Several structures of $V_{Ga}+n$H, $Bi_{Ga}+n$H and $H_i$ can be found in Supplementary Fig. S9.

In contract to $V_{Ga}$, hydrogen passivation does not remove the deep defect levels of $Bi_{Ga}$: both $Bi_{Ga}+$H and $Bi_{Ga}+2$H have two deep defect levels in the band gap, similar to $Bi_{Ga}$ (Fig. 5b). The binding energy of $Bi_{Ga}+$H is positive at $E_F < 0.29$ eV (Fig. 5d), indicating that H passivation cannot even be realized under such condition. For $Bi_{Ga}+2$H, the two hydrogen atoms form a $H_2$ molecule (Supplementary Fig. S9) rather than bind to $Bi_{Ga}$, and its defect formation energy curve resembles that of $Bi_{Ga}$ in shape (Fig. 5b).

Nevertheless, hydrogenation is anticipated to reduce the presence of defect $Bi_{Ga}$ by shifting up $E_F$. Figure 1a shows that $Bi_{Ga}$ exists primarily under p-type condition and its formation energy increases with $E_F$ until 0.81 eV, where $H_i$ is a donor (Fig. 5e). Therefore, $H_i$ will increase $E_F$ and exponentially decrease the equilibrium content of $Bi_{Ga}$ in the range of $E_F < 0.81$ eV. Consistent with these predictions, previous experiments indeed showed that hydrogenation reduced the DLTS signal of $As_{Ga}$ in GaAs.[51] Because of the similarity between $Bi_{Ga}$ and the other three primary defects, namely, $As_{Ga}$, $As_{Ga}+Bi_{As}$ and $Bi_{Ga}+Bi_{As}$, it is likely that hydrogenation would have similar effects on them too.

On the other hand, because $H_i$ introduces a defect level at 0.89 eV (Fig. 5e), which would negatively impact device performance, it is necessary to gradually increase the hydrogenation to identify the conditions that lead to the best device performance. Consistent with this observation, previous experiments did find that moderate hydrogenation prominently increased the photoluminescence intensity of GaAs but high hydrogen doses worsened it.[52]

**Influence of external chemical potentials.** Another method that can potentially reduce the unwanted defects is manipulating the external chemical potentials of Ga, As and Bi, which do not affect the positions of defect energy levels but strongly influence the defect formation energies. Figure 6a, 6b, and 6c show the formation energies of the point defects under As-rich, intermediate and Ga-rich conditions, respectively. The formation energies of defects involving the As site, namely $V_{As}$, $Ga_{As}$ and $Bi_{As}$, dramatically decrease when the chemical potential changes from more As-rich to more Ga-rich conditions, while the defects involving Ga-site, namely $V_{Ga}$, $As_{Ga}$ and $Bi_{Ga}$, sharply increase accordingly. Previous experiments indeed found that $V_{Ga}$ content increases with As vapor pressure.[33] Among the three conditions, the intermediate and Ga-rich conditions generate the fewest unwanted defects for the n- and p-type films, respectively. By contrast, the usual As-rich condition generates a significant number of unwanted defects for both types of doping.

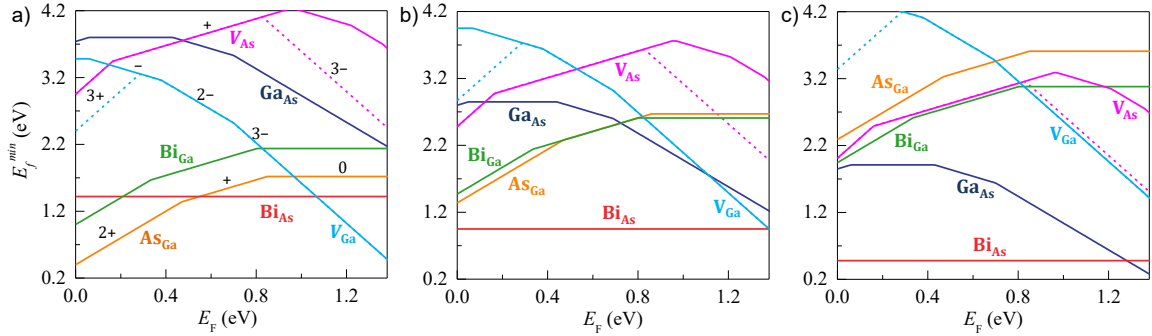

**Figure 6.** Lowest defect formation energy of point defects under the (**a**) As-rich, (**b**) intermediate and (**c**) Ga-rich conditions. The magenta and blue dashed branches correspond to the swapped structures of $V_{Ga}$ and $V_{As}$, respectively, as discussed in Table 2.

To avoid the As-rich growth condition, we suggest the As reactant being provided in a more precise way, e.g. using a pulsed As beam/precursor rather than filling the entire growth chamber with excessive As reactant, a method having also been proposed to increase the Bi ratio in GaAsBi.[12] Additionally, one might explore the pulsed laser deposition method,[53] which uses the plasma produced from GaAs bulk as reactant and thus creates chemical potentials close to the intermediate condition.

**SUMMARY**

In summary, we investigate the defect thermodynamics and Bi segregation in GaAsBi using DFT. We obtain defect energy levels that correspond reasonably well to the measured ones in GaAs and GaAsBi alloys, and provide valuable insight into the nature of previously observed defect energy levels in GaAsBi. We find that cation (anion) vacancy can change to anion (cation) vacancy at certain charge states, and this phenomenon exists in a number of semiconductors with low ionicity. Under the usual As-rich growth condition, $As_{Ga}$, $Bi_{Ga}$, $As_{Ga}+Bi_{As}$ and $Bi_{Ga}+Bi_{As}$ are the major minority-electron traps in p-type films, while $V_{Ga}$ and $V_{Ga}+Bi_{As}$ are the major minority-hole traps in n-type films. We predict that $V_{Ga}$ serves as the nuclei of the Bi-rich clusters and assists the diffusion of Bi defects. To reduce the deleterious effects of defects, we suggest using hydrogen passivation to decrease the minority-carrier traps and/or changing the growth to the Ga-rich or intermediate chemical potential conditions.

**ACKNOWLEDGMENTS**

This research was primarily supported by NSF through the University of Wisconsin Materials Research Science and Engineering Center (Grant No. DMR-1121288). Glen Jenness and Zhewen Song were supported by the NSF Software Infrastructure for Sustained Innovation (SI2) award No. 1148011. The authors gratefully acknowledge the use of computer clusters supported by NSF through the University of Wisconsin Materials Research Science and Engineering Center (Grant No. DMR-1121288). Computations in this work also benefited from the use of the Extreme Science and Engineering Discovery Environment (XSEDE), which is supported by National Science Foundation Grant No. ACI-1053575, the computing resources and assistance of the UW-Madison Center For High Throughput Computing (CHTC) in the Department of Computer Sciences, and the National Energy Research Scientific Computing Center (NERSC), a DOE Office of Science User Facility supported by the Office of Science of the U.S. Department of Energy under Contract No. DE-AC02-05CH11231.

**CONFLICT OF INTEREST**

The authors declare no conflict of interest.

# Supplemental Information for "Understanding and reducing deleterious defects in metastable alloy GaAsBi"

Guangfu Luo, Shujiang Yang, Glen Jenness, Zhewen Song, Thomas F. Kuech, Dane Morgan[*]
[*]E-mail:ddmorgan@wisc.edu

## I. Influence of hybrid functional and spin-orbit coupling on defect formation energy

Figure S1 compares the defect formation energy of $Bi_{As}$ and $Bi_{Ga}$ obtained using different methods. Two major features are observed. First, spin-orbit coupling (SOC) generally increases the defect formation energy of both defects, with an amplitude from 0.02 eV to 0.40 eV. The effects of SOC vary with charge state and therefore the defect energy levels change accordingly. Second, relative to local density approximation (LDA) functional, Heyd-Scuseria-Ernzerhof (HSE) functional[1] changes the magnitude of defect formation energies and in the case of $Bi_{Ga}$, the observed charge states and energy levels. Therefore, both hybrid functional and SOC are important to the correct descriptions of Bi defects.

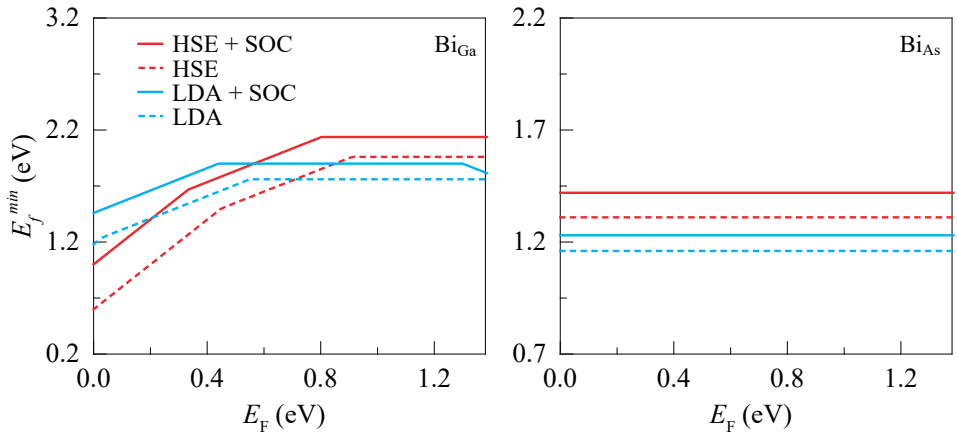

**Figure S1.** Lowest defect formation energy of (**Left**) $Bi_{Ga}$ and (**Right**) $Bi_{As}$ using four different methods: HSE functional and SOC (HSE+SOC), HSE functional, LDA functional and SOC (LDA+SOC), and LDA functional. Except functional and SOC, all other computational settings are the same for the four methods. Atomic structures are optimized using consistent parameters with the calculations of defect formation energy. LDA and LDA+SOC curves are obtained by plotting the formation energy of each charge state in the whole experimental band gap, ignoring the underestimated band gaps by LDA and LDA+SOC.

## II. Convergence tests of supercell size

Figure S2 shows the convergence tests of the formation energy of $V_{Ga}^{3-}$, $(V_{Ga}+Bi_{As})^{3-}$ and $(As_{Ga}+Bi_{As})^{2+}$ at the LDA level with and without the FNV correction. It is found that the errors of a 2 × 2 × 2 supercell (~64 atoms) without the FNV correction range from 0.5 eV to 0.9 eV relative to the dilute limits obtained with finite-size scaling approach.[2] In contrast, with the FNV correction, the average error of a 2 × 2 × 2 supercell is reduced to about 0.1 eV relative to either the FNV-corrected values of a 4 × 4 × 4 supercell (~512 atoms) or the dilute limits obtained with the finite-size scaling approach.

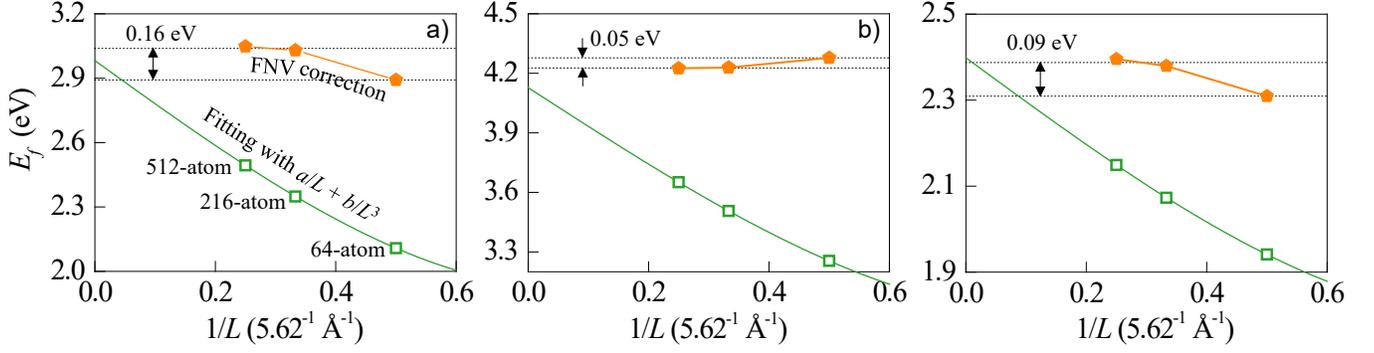

**Figure S2**. Supercell size tests of formation energy using LDA functional at $E_F = E_{VBM}$ for (**a**) $V_{Ga}^{3-}$, (**b**) $(V_{Ga}+Bi_{As})^{3-}$ and (**c**) $(As_{Ga}+Bi_{As})^{2+}$. $L$ is the lattice length of supercell.

### III. Comparison between experiments and early theoretical predictions

Figure S3 shows that previous theoretical predictions on defect energy levels in GaAs agree poorly with experiments. Although some theoretical points are close to the experimental values, other theoretical values from the same reference deviate significantly from the experimental values.

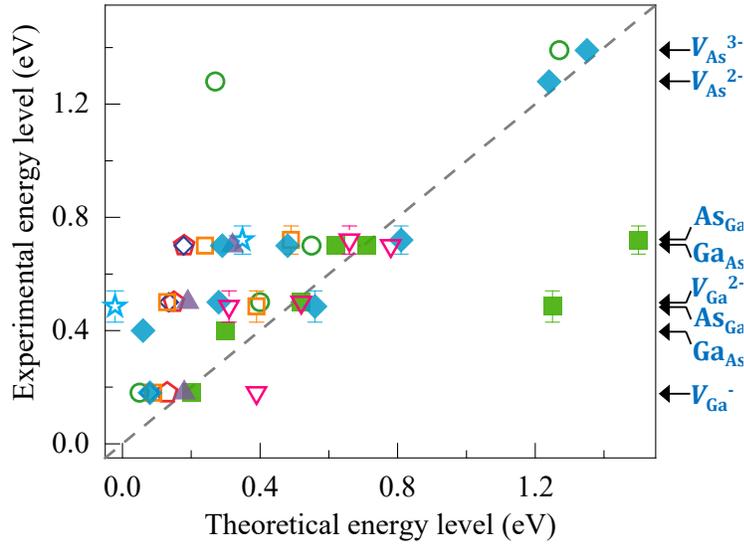

**Figure S3.** Comparison of defect energy levels between experiments and previous theoretical predictions. Experimental data: $V_{Ga}$ from Ref. 3, 4, $V_{As}$ from Ref. 5, 6, $As_{Ga}$ from Ref. 7-11 and $Ga_{As}$ from Ref. 12-14. Theoretical data: ■ from Ref. 15, ■ from Ref. 16, ▲ from Ref. 17, □ from Ref. 18, △ from Ref. 19, ⬠ from Ref. 20, ○ from Ref. 21, ▢ from Ref. 22 and ★ from Ref. 23. The error bars of the experimental data represent the range of experimental values.

### IV. Change of anion vacancy to cation vacancy in zincblende binary compounds

Table S1 shows that the change of anion vacancy to cation vacancy is thermodynamically favorable for most zincblende binary compounds that we examined, and the compounds with lower ionicity tend to undergo such change over a larger number of charge states. This trend is expected because more ionic systems will typically have more unstable antisite energies due to the large chemistry difference between the atoms involved.

**Table S1.** Ionicity of twelve zincblende binary compounds and the charge states, at which the swapped structure is more stable than the anion vacancy.

| Compound $AB$ | Ionicity[24] | charge state of $V_B \rightarrow V_A + A_B$ |
|---|---|---|
| ZnO | 0.616 | Does not occur |
| GaN | 0.500 | 3- |
| InP | 0.421 | 3- |
| InAs | 0.357 | 3- |
| GaP | 0.327 | 2-, 3- |
| InSb | 0.321 | 2-, 3- |
| AlP | 0.307 | 2-, 3- |
| GaAs | 0.310 | 2-, 3- |
| AlAs | 0.274 | 2-, 3- |
| GaSb | 0.261 | 1-, 2-, 3- |
| AlSb | 0.250 | 1-, 2-, 3- |
| SiC | 0.177 | 3-, 4- |

## V. Estimation on the contents of Bi defects under non-equilibrium conditions

In order to estimate the contents of Bi defects under non-equilibrium conditions, which usually lead to much higher total Bi contents (>1%) than the equilibrium prediction, we utilize the defect formation energies in Fig. 1, fix the total Bi content to a typical experimental value of 1%, and allow the Bi chemical potential $\mu_{Bi}$ to vary as necessary, as defined by Eqn. S1,

$$\sum_i N_i e^{-E_f^{min}(d_i, E_F, \mu_{Bi})/(k_B T)} = 0.01 \tag{S1}$$

where $d_i$ is a Bi defect and $N_i$ is the number of Bi atom per unit cell of GaAs bulk for a defect $d_i$. Figure S4 shows the defect formation energies and mole fractions of Bi defects under the condition of Eqn. S1. It is found that there are five major Bi defects throughout the range of $E_F$, namely, $Bi_{Ga}+Bi_{As}$ and $Bi_{Ga}$ under heavy to medium p-type doping condition, $Bi_{As}$ and $Bi_{As}+Bi_{As}$ under light p-type doping to medium n-type doping condition, and $V_{Ga}+Bi_{As}$ and $Bi_{As}$ under heavy n-type doping condition. The most significant change relative to the equilibrium predictions (Fig. 1) is a dramatic increase in the relative fractions of $Bi_{Ga}+Bi_{As}$ and $Bi_{As}+Bi_{As}$.

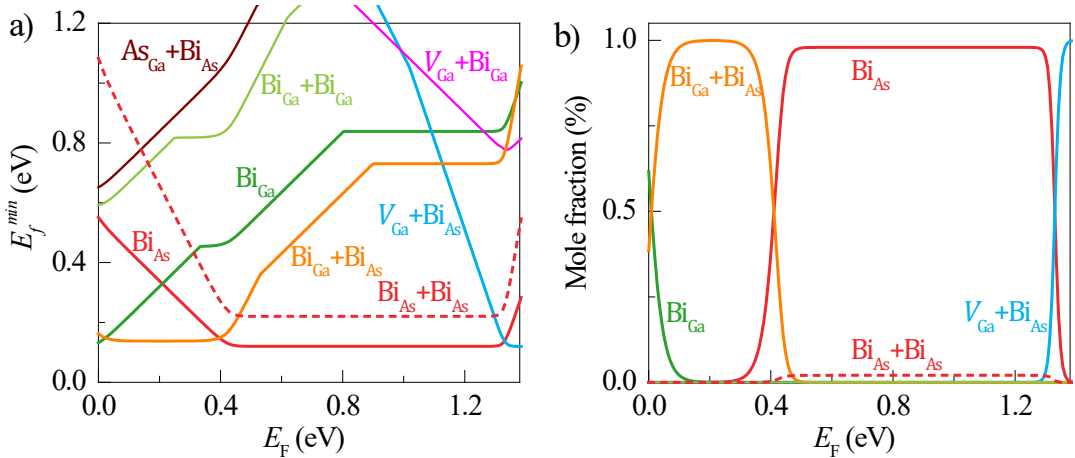

**Figure S4.** Estimated (a) defect formation energies and (b) contents of Bi defects under non-equilibrium condition as defined by Eqn. S1. Temperature is set to 300 K.

## VI. Minority-carrier traps with consideration of only equilibrium defect formation energies

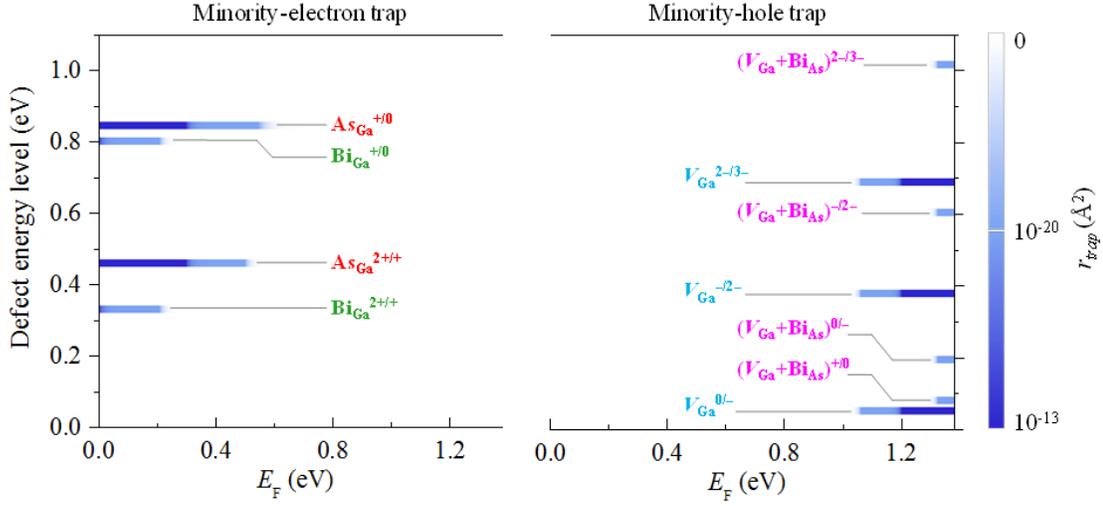

**Figure S5.** Trapping strength as a function of Fermi energy for (**Left**) minority-electron and (**Right**) minority-hole traps induced by the energy levels of all examined point and pair defects. Data is calculated with the equilibrium formation energies from Fig. 1. Energy levels of $Bi_{Ga}+Bi_{As}$ (see Fig. 3), which may be present under non-equilibrium conditions (see Fig. S4), are absent because of its high formation energy at equilibrium.

## VII. Defect formation energy and binding energy of $V_{As}+nBi_{As}$ and $V_{Ga}+nBi_{As}$

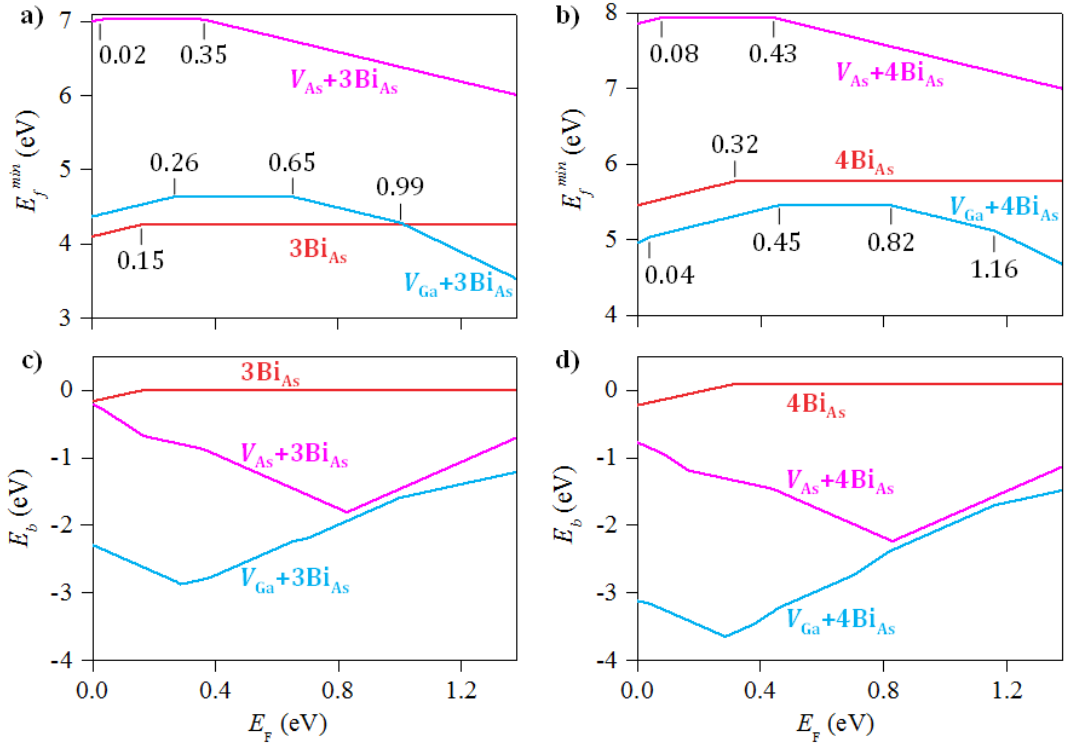

**Figure S6.** Lowest defect formation energies of (**a**) $V_{Ga}+3Bi_{As}$ and (**b**) $V_{Ga}+4Bi_{As}$, and binding energies of (**c**) $V_{Ga}+3Bi_{As}$ and (**d**) $V_{Ga}+4Bi_{As}$.

## VIII. Diffusion of $Bi_{Ga}$ mediated by $V_{Ga}$ hopping

S4

To evaluate the diffusion coefficient and diffusion length of $Bi_{Ga}$ mediated by $V_{Ga}$ hopping, we utilize the widely-used five-frequency model.[25, 26] This model requires the frequencies of five processes, namely, $V_{Ga}$ hop in GaAs bulk (process 0), $V_{Ga}$ rotation-hop around $Bi_{Ga}$ (process 1), $Bi_{Ga}$ exchange with $V_{Ga}$ (process 2), dissociation of $V_{Ga}$ and $Bi_{Ga}$ (process 3) and association of $V_{Ga}$ and $Bi_{Ga}$ (process 4), as shown in Fig. S7. Our calculations based on the climbing nudged elastic band method[27] find that the energy barriers of processes 0 – 4 are 2.16, 2.47, 1.79, 2.54 and 1.98 eV, respectively. Note that we consider the As-rich and $n$-type doping conditions, where the process 0 has the 3- charge state and the processes 1– 4 are in the charge state 1-.

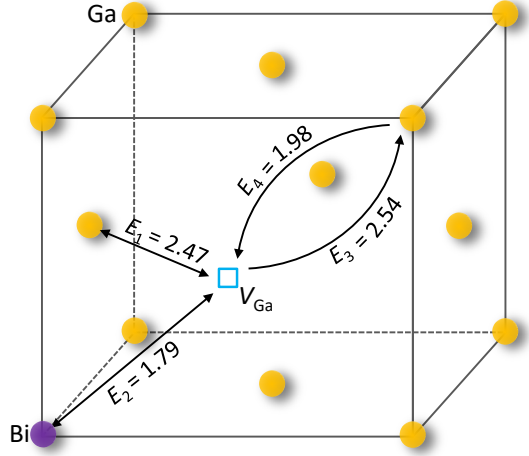

**Figure S7.** Four hopping barriers (in unit of eV), $E_1 - E_4$, used in the five-frequency model for the $V_{Ga}$-mediated $Bi_{Ga}$ diffusion. The As atoms are hidden for easy visualization.

Then these energy barriers are used to calculate the corresponding frequencies according to $\omega_i = \upsilon_i e^{-E_i/(k_B T)}$, where $E_i$ is hopping barrier and $\upsilon_i$ are the attempt frequencies. We assume all the $\upsilon_i$ are equal to a single $\upsilon$ and note that they cancel out in the final equations (Eqn. S2–S4), except that in $D_0$. The diffusion coefficient of $Bi_{Ga}$, $D$, is determined by the five frequencies according to Eqn. S2–S4.[25, 26, 28]

$$D = \frac{f_2}{f_0}\frac{\omega_2}{\omega_0}\frac{\omega_4}{\omega_3} D_0; \quad D_0 = \xi e^{-(E_f + E_0)/(k_B T)} \tag{S2}$$

$$f_2 = \frac{1 + \frac{7}{2} F(\omega_4/\omega_0) \times (\omega_3/\omega_1)}{1 + \frac{7}{2} F(\omega_4/\omega_0) \times (\omega_3/\omega_1) + (\omega_2/\omega_1)} \tag{S3}$$

$$F(x) = 1 - \frac{10x^4 + 180.5x^3 + 927x^2 + 1341x}{7(2x^4 + 40.2x^3 + 254x^2 + 597x + 436)} \tag{S4}$$

Here the correlation factor $f_0$ equals 0.7815 for the FCC lattice. The Ga self-diffusion coefficient $D_0$ depends on formation energy $E_f$, which equals 4.62eV–3$E_F$ (see Fig. 1a), and migration barrier $E_0$ of $V_{Ga}^{3-}$, which equals 2.16 eV. $D_0$ also depends on the coefficient $\xi$, which includes information of $f_0$, lattice constant, attempt frequency, formation entropy and migration entropy. Here we adopt $\xi = 4.3 \times 10^{17}$ Å$^2$/sec from a previous isotope experiment on Ga self-diffusion in intrinsic GaAs, which was measured in the temperature range of 800–1225 °C under As-rich condition.[29] Note that our $E_f + E_0$ equals 4.35 eV for intrinsic GaAs when the $E_F$ is ~0.1 eV above gap center in the temperature range of 800–1225 °C,[30] and this barrier agrees well with the experimental value of 4.24 eV.[29] Finally, the diffusion length, $l$, in time $t$ is determined according to Eqn. S5.

S5

$$l = \sqrt{6\,D\,t} \tag{S5}$$

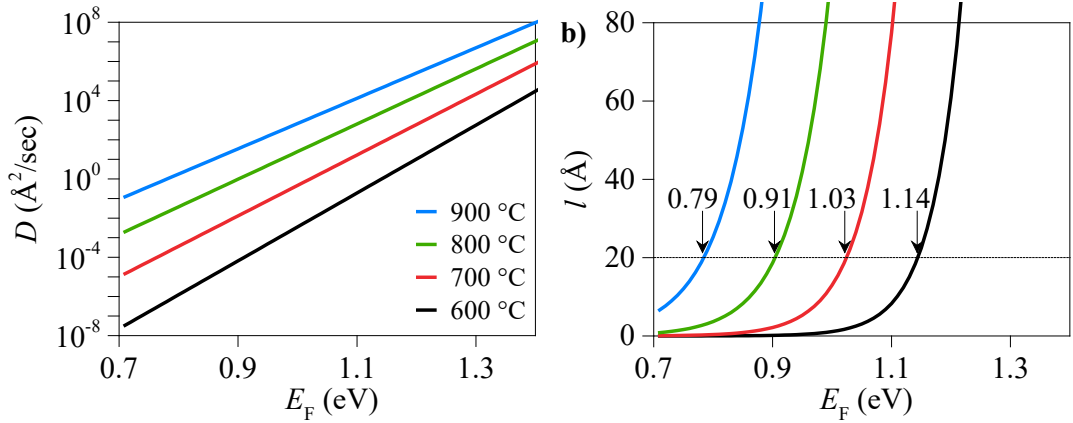

**Figure S8.** (**a**) Diffusion coefficient of $V_{Ga}$–mediated $Bi_{Ga}$ and (**b**) diffusion length in 60 seconds as a function of $E_F$ at different annealing temperatures.

To quantify the diffusivity and diffusion length of $V_{Ga}$-mediated $Bi_{Ga}$, we consider the fairly typical thermal annealing conditions of GaAsBi, namely, 60–120 seconds annealing in the temperature range of 600–800 °C.[31] Figure S8 plots the diffusivity and diffusion length of $Bi_{Ga}$ as a function of $E_F$ under different annealing conditions. The diffusion length is over 20 Å for a 60-second annealing at a temperature of 600, 700, 800 and 900 °C, when $E_F$ is over 1.14, 1.03, 0.91 and 0.79 eV, respectively. In comparison, for a GaAsBi with 1.0% to 3.0% uniformly distributed Bi atoms, the average distance between two neighbor Bi atoms is only 16 to 11 Å. Therefore, under typical annealing conditions, the $V_{Ga}$–mediated $Bi_{Ga}$ diffusion can be fast enough to accumulate and form Bi-rich clusters.

## IX. Atomic structures of $V_{Ga}+n$H, $Bi_{Ga}+n$H and $H_i$

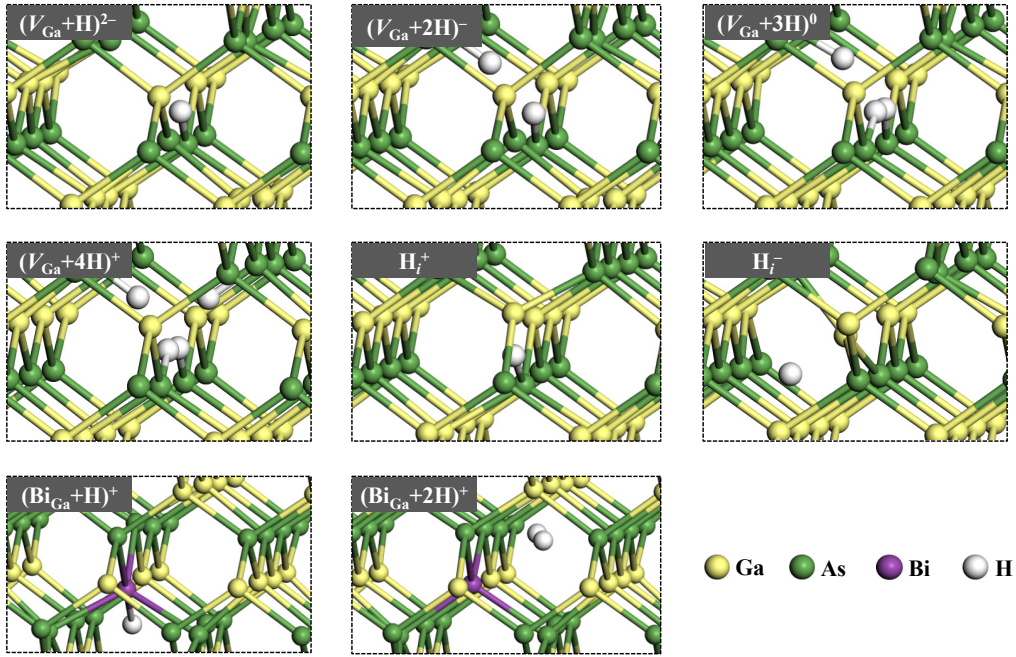

**Figure S9.** Several atomic structures of $V_{Ga}+n$H, $Bi_{Ga}+n$H and $H_i$.

S6